\documentclass[onecollarge]{svjour2}       
\smartqed  
\usepackage{psfrag,graphicx}
\journalname{Journal of Statistical Physics}
\begin{document}

\title{document}
\title{Lyapunov exponent for small particles in smooth one-dimensional flows}

\author{Michael Wilkinson}

\institute{Michael Wilkinson \at
              Department of Mathematics and Statistics, \\
              The Open University,
              Walton Hall, \\
              Milton Keynes, MK7 6AA, \\
              England.\\
              \email{m.wilkinson@open.ac.uk}      \\
}

\date{Received: date / Accepted: date}

\maketitle

\begin{abstract}

This paper discusses the Lyapunov exponent $\lambda$ for small 
particles in a spatially and temporally smooth flow in one dimension. 
Using a plausible model for the statistics of the velocity gradient
in the vicinity of a particle, the Lyapunov exponent is obtained as a 
series expansion in the Stokes number, ${\rm St}$, which is a dimensionless 
measure of the importance of inertial effects. The approach described here 
can be extended to calculations of the Lyapunov exponents and of the 
correlation dimension for inertial particles in higher dimensions. 
It is also shown that there is correction to this theory which arises because 
the particles do not sample the velocity field ergodically. Using this 
non-ergodic correction, it is found that (contrary to expectations) the 
first order term in the expansion of $\lambda$ does not vanish. 


\end{abstract}

\maketitle

\section {Introduction}
\label{sec: 1}

It is known that small particles suspended in complex flows, such as 
fully-developed turbulence, can cluster together \cite{Max87}, so that 
they accumulate on a fractal attractor \cite{Som+93}. In cases where the 
flow has a compressible component, it is also possible for the trajectories 
of particles to coalesce \cite{Deu85}. These effects have been analysed by 
calculating the Lyapunov exponents of the particle trajectories (the role 
of Lyapunov exponents in characterising dynamical systems is discussed in 
\cite{Ott02}). The leading Lyapunov exponent $\lambda$ is the average of the 
logarithmic rate of divergence of particles with an infinitesimal separation 
$\delta x(t)$: 
\begin{equation}
\label{eq: 1.1}
\lambda=\lim_{t\to \infty}\frac{1}{t}\left\langle{\rm ln}\left[\frac{\delta x(t)}{\delta x(0)}\right]\right\rangle
\end{equation}
where $\langle X\rangle$ denotes the expectation value of $X$. If the 
leading Lyapunov exponent is negative, the particles aggregate together 
(the path coalescence effect) \cite{Wil+03}. In cases where there is 
clustering onto a fractal attractor rather than aggregation, the dimension 
of the attractor can be estimated \cite{Bec+06} from the full set of 
Lyapunov exponents of the particle trajectories by means of the Lyapunov 
dimension, given by the Kaplan-Yorke formula \cite{Kap+79}. The clustering 
effect may play an important role in explaining the growth of rain drops 
\cite{Sha03} and of planetesimals \cite{Cuz+01}, and deserves to be thoroughly 
understood. 

Substantial progress has been made in analysing these clustering effects for 
a model with a random velocity field, in the limit where the correlation time of the velocity 
field is vanishingly small. All three Lyapunov exponents for the suspended 
particles were obtained as power series in a dimensionless parameter \cite{Wil+07}. 
The methods used in that work involve a mapping to a Fokker-Planck equation, and 
appear to be restricted to the case of a vanishing correlation time, where the system 
is described by a Markovian stochastic process. This paper shows how the analysis 
developed in \cite{Wil+07} may be extended to models where the velocity field has 
a finite correlation time, using a result from a companion paper, 
\cite{Wil09}. This allows the theoretical analysis of clustering to be extended to 
physically realistic model velocity fields. 

It will be shown that there is an unexpected subtlety in the application of the finite 
correlation time model, which may complicate its application to models of turbulent 
velocity fields. For this reason, the problem is explored here in its simplest form, 
involving a one-dimensional flow, so that the difficulties can be explored in the
simplest context.

In order to analyse the clustering effect it is necessary to consider the dimensionless 
parameters of this problem. The Stokes number, a dimensionless measure of the inertia of 
the particles, is defined by 
\begin{equation}
\label{eq: 1.2}
{\rm St}=\frac{1}{\gamma \tau}
\end{equation}
where $\tau$ is the correlation time for the fluctuations of the velocity gradient 
of the fluid, and $\gamma$ is the rate at which particle motion relative to the 
fluid is damped. The statistics of the velocity field are described 
by a dimensionless parameter known as the Kubo number:
\begin{equation}
\label{eq: 1.3}
{\rm Ku}=\frac{u_0\tau}{\xi}
\end{equation}
where $u_0$ is a typical scale size of the velocity field and $\xi$ is a 
correlation length. In \cite{Wil+07} it is argued that for fully-developed 
turbulence the relevant length, time and velocity scales are those corresponding 
to the Kolmogorov microscales of the velocity field, and that dimensional 
considerations imply that the Kubo number is of order unity
for turbulent flows. However, the Kubo number will remain as a parameter in any random flow 
model for turbulence. It is desirable to understand how the Lyapunov exponent
varies as a function of both the Stokes number and the Kubo, so that the analytical 
properties of the approximations can be thoroughly understood. In section \ref{sec: 3}, 
the variables ${\rm St}$ and ${\rm Ku}$ will be replaced by variables $\omega$ and $\kappa$, which 
are dimensionally equivalent but which are precisely defined. 

The model with vanishing correlation time which is analysed in \cite{Wil+07} 
corresponds to considering the limit as ${\rm Ku}\to 0$. The application of models
with with zero correlation time has been criticised \cite{Bec+07}, because it is argued
that that such models fail to get the leading order terms of series expansions 
for the Lyapunov exponents correct. In particular, it has been argued \cite{Bec+07} that the 
difference between the fractal dimension $D$ of the measure upon which particles 
cluster and the dimension $d$ of the coordinate space is of second order in the Stokes parameter, 
that is $d-D=O({\rm St}^2)$, whereas for the zero correlation time model it 
is found that $D-d=O({\rm St})$ \cite{Wil+07}. Numerical evidence for the prediction
that $d-D=O({\rm St}^2)$ is presented 
by Bec \cite{Bec03}, who credits Balkovsky {\sl et al} \cite{Bal+01} with 
the theoretical basis for this result. This discrepancy is addressed by the model which 
is developed here, and it emerges that these issues can be quite subtle. 
For the one-dimensional model with zero correlation time it is known \cite{Wil+03} that 
$\lambda/\gamma=O({\rm St})$, and here it is shown that a simple 
application of the model with finite correlation time indicates that the 
leading order dependence of $\lambda/\gamma$ is $O({\rm St}^2)$, 
in line with the expectation that the finite correlation time model should 
have a different leading-order behaviour. A more careful analysis, however, 
leads to a surprising conclusion. The model with finite correlation time has 
the same leading order behaviour as the zero correlation time model, namely 
$\lambda/\gamma=O({\rm St})$. Moreover, the prefactor is the same.

Section \ref{sec: 2} explains how the calculation of Lyapunov exponents is 
related to the problem discussed in the companion paper \cite{Wil09}, which
analyses the statistics of a class of stochastic differential equations which
are driven by noise signal which has a finite correlation time. Section \ref{sec: 3} 
shows how the series expansion for the expectation value derived in \cite{Wil09} 
is used to produce an estimate for the Lyapuonov exponent, which satisfies 
$\lambda/\gamma=O({\rm St}^2)$. Section \ref{sec: 4} shows that 
this simple estimate is based upon an unjustified assumption, namely that 
the positions of the particles sample the velocity gradient field ergodically.
The leading order non-ergodic correction is computed, and it is shown that the 
correct leading order behaviour is $\lambda/\gamma=O({\rm St})$. 
A different example of the importance of non-ergodic effects in the
dynamics of small particles in a random velocity field was commented
upon in \cite{Meh+09}.
Section \ref{sec: 5} is a summary of the conclusions, and points the way 
to applications in the systematic analysis of particle clustering in realistic 
two- and three-dimensional flows.

\section{A model for the Lyapunov exponent of particles in a random flow}
\label{sec: 2}

This section discusses the connection between the Lyapunov exponent and 
the equation of motion (\ref{eq: 1.1}) of the companion paper, \cite{Wil09}. Equations (\ref{eq: 2.1}) 
to (\ref{eq: 2.6}) follow arguments in \cite{Wil+03} closely, but 
this material is included for the convenience 
of the reader.

Consider a one-dimensional model for particles embedded in a fluid with 
velocity $u(x,t)$. It is assumed that the motion of the particle relative to the fluid 
experiences a viscous damping force which is proportional to the velocity 
of the particle relative to the fluid. The equation of motion for a particle 
with position $x$ and velocity $v$ is therefore
\begin{equation}
\label{eq: 2.1}
\dot x=v\ ,\ \ \ \dot v=-\gamma [v-u(x,t)]
\ .
\end{equation}
The Lyapunov exponent characterises the time dependence of the infinitesimal 
separation $\delta x$ of a trajectory from a reference trajectory. To analyse 
this quantity, linearise the equation of motion to obtain
\begin{equation}
\label{eq: 2.2}
\delta \dot v=\delta x\ ,\ \ \ \delta \dot v=-\gamma[\delta v-Y(t)\delta x]
\end{equation}
where
\begin{equation}
\label{eq: 2.3}
Y(t)=\frac{\partial u}{\partial x}(x(t),t)
\end{equation}
is the velocity gradient at the current position of the particle. Now define 
a variable $X$:
\begin{equation}
\label{eq: 2.4}
X=\frac{\delta v}{\delta x}
\ .
\end{equation}
This has a simple relation to the Lyapunov exponent:
\begin{equation}
\label{eq: 2.5}
\lambda=\langle X \rangle
\end{equation}
and the  equation of motion for $X$ is 
\begin{equation}
\label{eq: 2.6}
\dot X=-\gamma X-X^2+\gamma Y(t)
\ .
\end{equation}
Thus the task of calculating $\lambda$ resolves into calculating the 
expectation value of a variable $X$ which satisfies equation (\ref{eq: 2.6}). 

In turbulent or other unsteady, complex flows the velocity gradient 
$Y(t)=\partial u(x(t),t)/\partial x$ can be modelled as a stochastic variable. 
With this interpretation, equation (\ref{eq: 2.6}) is equivalent to 
the stochastic differential equation (\ref{eq: 1.1}) of the companion paper
\cite{Wil09}. The velocity gradient has a finite correlation time, with a 
correlation function which depends upon the details of the flow. If the fluid 
velocity field $u(x,t)$ is statistically homogeneous in space and time, it 
seems reasonable to model the fluctuations of the velocity gradient by a 
random variable which has mean value equal to zero. The companion paper 
discussed the case where the random variable $Y(t)$ has a finite correlation 
time, by considering the following model for $Y(t)$:
\begin{equation}
\label{eq: 2.7}
\dot Y=-\frac{1}{\tau} Y+\eta(t)
\end{equation}
with
\begin{equation}
\label{eq: 2.8}
\langle \eta(t)\rangle=0 \ ,\ \ \ \langle \eta(t_1)\eta(t_2)\rangle=2D\delta(t_1-t_2)
\ .
\end{equation}
This is an Ornstein-Uhlenbeck process \cite{Uhl+30}, discussed in \cite{vKa81}. 
The correlation function of this process is 
\begin{equation}
\label{eq: 2.9}
\langle Y(t_1)Y(t_2)\rangle=D\tau \exp(-\vert t_1-t_2\vert/\tau)
\end{equation}
which implies that 
\begin{equation}
\label{eq: 2.10}
D=\frac{1}{2\tau^2}\int_{-\infty}^\infty {\rm d}t\ \langle Y(t)Y(0)\rangle
\end{equation}
so that the diffusion coefficient in (\ref{eq: 2.8}) is related to the 
correlation function of $Y(t)$ by (\ref{eq: 2.10}).

\section{A series expansion for the Lyapunov exponent}
\label{sec: 3}

In the following equations (\ref{eq: 2.6}), (\ref{eq: 2.7}) and 
(\ref{eq: 2.8}) will be transformed into a dimensionless form which is precisely the 
same as equations (4), (5) of the companion paper, \cite{Wil09}. The results developed 
there will be used to obtain a series expansion of the Lyapunov exponent, 
$\lambda=\langle X\rangle$. It is convenient 
to transform to a new dimensionless time variable, $t'=\gamma t$. And the 
variables $X$, $Y$ are related to dimensionless variables by writing 
$X=\alpha X'$, $Y=\alpha Y'$, where the quantity $\alpha$ has dimensions of 
inverse time. By writing $\alpha=\tau\sqrt{D\gamma}$ the equations of motion 
(\ref{eq: 2.6}), (\ref{eq: 2.7}) can be written in the form
\begin{eqnarray}
\label{eq: 3.1}
\frac{{\rm d}X'}{{\rm d}t'}&=&Y'-X'-\epsilon X'^2
\nonumber \\
\frac{{\rm d}Y'}{{\rm d}t'}&=&-\omega Y'+\omega\zeta(t')
\end{eqnarray}
where $\zeta(t')$ is a white noise with statistics
\begin{equation}
\label{eq: 3.2}
\langle \zeta(t')\rangle=0\ ,\ \ \ \langle \zeta(t'_1)\zeta(t'_2)\rangle=2\delta (t'_1-t'_2)
\ .
\end{equation}
Equations (\ref{eq: 3.1}), (\ref{eq: 3.2}) are precisely the system which is 
analysed in the companion paper, \cite{Wil09}. The parameters $\epsilon$ and 
$\omega$ in (\ref{eq: 3.1}), and an additional parameter $\kappa$, are defined by
\begin{equation}
\label{eq: 3.3}
\epsilon=\tau\sqrt{\frac{D}{\gamma}}\ ,\ \ \ \omega=\frac{1}{\gamma \tau}
\ ,\ \ \ \kappa=\sqrt{D\tau^3}\ .
\end{equation}
The dimensionless parameters $\kappa$ and $\omega$ are precisely 
defined realisations of the dimensionless parameters introduced in section 
\ref{sec: 1}, that is
\begin{equation}
\label{eq: 3.4}
{\rm St}=\omega\ ,\ \ \ {\rm Ku}=\kappa\ ,\ \ \ \epsilon ={\rm Ku}\,{\rm St}^{1/2} 
\ .
\end{equation}
The Lyapunov exponent is 
\begin{equation}
\label{eq: 3.5}
\lambda=\alpha \langle X'\rangle=\gamma \epsilon \langle X' \rangle\ .
\end{equation}
The companion paper gives a series expansion for the expectation value 
$\langle X'\rangle$. Using (\ref{eq: 3.4}) to express $\epsilon$ in terms 
of $\kappa$ and $\omega$ and using equation (30) of \cite{Wil09} gives
\begin{equation}
\label{eq: 3.6}
\lambda=\gamma \kappa^2 \omega^2\left[\frac{1}{1+\omega}+C_1(\omega)\kappa^2\omega^2+C_2(\omega)\kappa^4\omega^4+\ldots\right]
\end{equation}
where the $C_j(\omega)$ are ratios of polynomials with the same degree, with 
$C_2(0)=15\, ,C_3(0)=630\, \ldots$. The leading order behaviour of the 
zero correlation time model is 
$\lambda \sim \gamma \epsilon^2=\gamma \kappa^2\omega$ \cite{Wil+03}. Thus 
equation (\ref{eq: 3.6}) is in agreement with the expectation that for a 
velocity field with finite correlation time, the leading order behaviour is 
different. It also has the satisfying feature that it 
is based upon a model which yields all of the coefficients of the series 
expansion exactly by using the operator algebra discussed in the companion 
paper. The technique appears to have a straightforward extension to multi-dimensional cases, and 
can also be adapted to provide the full series expansion of the subdominant 
Lyapunov exponents \cite{Wil+07} and the 
correlation dimension \cite{Wil+09}.

This is however a deficiency in the calculation, which will be discussed in 
the next section. 

\section{Non-ergodic correction to the velocity statistics}
\label{sec: 4}

Note that the model introduced in section \ref{sec: 2} uses the seemingly 
uncontroversial assumption that the mean value of the velocity gradient 
$Y(t)$ is zero. For a velocity field $u(x,t)$ which is statistically 
homogeneous as a function of $x$, the expectation value of $\partial u/\partial x$
must certainly vanish. It is, however, possible for $\langle Y(t)\rangle$ to be 
non-zero because $Y(t)$ is the velocity gradient evaluated at the current 
position of the particle, $x(t)$. Because $x(t)$ is itself dependent upon 
the history of the velocity field, $Y(t)$ may not represent an ergodic 
sampling of this field.

This non-ergodic correction may, in most circumstances, represent a small refinement
of the theory. In the overdamped limit (as $\omega \to 0$) it could, however, be of critical
importance in determining the leading order behaviour of $\lambda$. In the limit as
$\omega \to 0$, the Ornstein-Uhlenbeck noise term $Y'$ in equation (\ref{eq: 3.1})
varies very slowly. The stochastic variable $X'$ must therefore also vary very slowly,
and the time-derivative in (\ref{eq: 3.1}) can therefore be neglected. In the 
overdamped limit, the equation of motion for $X'$ can therefore be approximated by
\begin{equation}
\label{eq: 4.1}
Y'-X'-\epsilon X'^2=0
\ .
\end{equation}
This is a quadratic equation determining $X'$ as a function of $Y'$.
In the overdamped limit (where $\omega\to 0$ with $\kappa$ fixed), 
$\epsilon=\kappa \sqrt{\omega}\to 0$. The solution of this quadratic equation 
is approximated by $X'=Y'-\epsilon Y'^2+O(\epsilon^2)$. Taking expectation values, and 
using (\ref{eq: 3.5}), this implies that the Lyapunov exponent is
\begin{equation}
\label{eq: 4.2}
\lambda=\gamma \epsilon \left[\langle Y' \rangle-\epsilon \langle Y'^2\rangle \right]+O(\epsilon^3)
\ .
\end{equation}
For the Ornstein-Uhlenbeck noise process in (\ref{eq: 3.1}), 
$\langle Y'\rangle=0$ and $\langle Y'^2\rangle=\omega$, so that the 
approximate equation (\ref{eq: 4.2}) gives 
$\lambda=\gamma\epsilon^2\omega=\gamma \kappa^2\omega^2$, a result in precise agreement with
the leading order term in (\ref{eq: 3.5}). If, however, a more refined approximation 
shows that $\langle Y(t)\rangle$ is non-zero, it may be discovered that the dominant 
contribution to $\lambda/\gamma$ does, in fact, come from a term which is linear 
in the Stokes number $\omega$.

In principle it is possible to expand $\langle Y\rangle$ as a series in $\omega$ and $\kappa$. 
Only the leading term of this expansion is considered here. 
In the limit where $\omega \ll 1$ a particle is simply advected by the flow
and the equation of motion may be approximated by the advection 
equation, $\dot x=u(x(t),t)$. The position of the particle is then
\begin{equation}
\label{eq: 4.3}
x(t)=x(0)+\int_0^t{\rm d}t'\ u(x(t'),t')+O(\omega)
\ .
\end{equation}
Now use this approximation to calculate the time-average of $Y(t)$, that is
\begin{equation}
\label{eq: 4.4}
\langle Y(t)\rangle=\lim_{T\to \infty} \biggl[
{1\over T}\int_0^T{\rm d}t\ \frac{\partial u}{\partial x}(x(t),t)\biggr]
\ .
\end{equation}
Using equation (\ref{eq: 4.3}),
\begin{equation}
\label{eq: 4.5}
\int_0^T{\rm d}t\ \frac{\partial u}{\partial x}(x(t),t)=\int_0^T {\rm d}t\ 
\biggl[{\partial u\over{\partial x}}(x(0),t)+{\partial^2 u\over{\partial x^2}}(x(0),t)
\int_0^t{\rm d}t'\ u(x(0),t')\biggr]+\ldots \ .
\end{equation}
Averaging this, using translational invariance and assuming $T\gg \tau $ gives
\begin{eqnarray}
\label{eq: 4.6}
\langle Y(t)\rangle &=&\lim_{T\to \infty}\frac{1}{T}\int_0^T{\rm d t}\int_0^t{\rm d}t'\ 
\langle \partial^2_x u(0,t) u(x,t')\rangle + \ldots
\nonumber \\
&=&\frac{1}{2}\int_{-\infty}^\infty {\rm d}t\ 
\langle \partial^2_x u(0,t)\,u(0,0)\rangle +\ldots
\ .
\end{eqnarray}
The correlation function in (\ref{eq: 4.6}) is easily related to that in 
(\ref{eq: 2.10}). To this end define correlation functions $C_0(t)$, 
$C_1(t)$ and $C_2(t)$ as follows: 
\begin{equation}
\label{eq: 4.7}
C_0(t)=\left\langle \frac{\partial}{\partial x}\left[\frac{\partial u}{\partial x}(x,t)u(x,0)\right]\right\rangle
, \ 
C_1(t)=\left\langle \frac{\partial^2 u}{\partial x^2}(x,t)u(x,0)\right\rangle
, \ 
C_2(t)=\left\langle \frac{\partial u}{\partial x}(x,t)\frac{\partial u}{\partial x}(x,0)\right\rangle
\ .
\end{equation}
Note that $C_0(t)=C_1(t)+C_2(t)$. Consider the evaluation of $C_0(t)$ by averaging
over the spatial coordinate: because the quantity being averaged is a spatial 
derivative,
\begin{equation}
\label{eq: 4.8}
C_0(t)=\lim_{L\to \infty}\frac{1}{L}\int_0^L {\rm d}x\ \frac{\partial}{\partial x}
\left[\frac{\partial u}{\partial x}(x,t)u(x,0)\right]=0
\end{equation}
implying that $C_1(t)=-C_2(t)$. Now return to the evaluation of $\langle Y(t)\rangle$
using (\ref{eq: 4.6}). Recalling the definition of $D$ in (\ref{eq: 2.10}),
\begin{equation}
\label{eq: 4.9}
\langle Y(t)\rangle=\frac{1}{2}\int_{-\infty}^\infty {\rm d}t\ C_1(t)
=-\frac{1}{2}\int_{-\infty}^\infty {\rm d}t\ C_2(t)
=-\tau^2 D \ .
\end{equation}
From this it follows that $\langle Y'\rangle=\tau\sqrt{D/\gamma}=\epsilon$, and hence, using
(\ref{eq: 4.2}), one obtains the following estimate for the Lyapunov exponent 
\begin{equation}
\label{eq: 4.10}
\lambda=-\gamma \epsilon^2=-\gamma \kappa^2 \omega\ .
\end{equation}
This is exactly the form obtained using the zero correlation time model, discussed in \cite{Wil+03}.

\section{Discussion}
\label{sec: 5}

This paper has discussed a new approach to extending the theoretical 
understanding of the dynamics of particles suspended in random flows, 
through making series expansions of the Lyapunov exponent in powers 
of the Stokes number. Until now, this has only been possible for models 
of the fluid velocity field with zero correlation time. This paper has 
shown how the techniques for obtaining the series expansion can be extended 
to a class of models for flows with a finite correlation time.

It has been questioned whether the leading order behaviour of these series is 
correct for the zero correlation time model where, $\kappa\to 0$. For 
example, it has been proposed that the difference between the fractal dimension of the 
attractor and the space dimension is $O({\rm St}^2)$ for real flows, whereas it is found to 
be $O({\rm St})$ for the zero correlation time model. A direct application of the finite
correlation time model, as considered in section \ref{sec: 4}, appears to be 
consistent with this expectation, in that the leading order dependence of 
$\lambda/\gamma$ appears to be $O({\rm St}^2)$ for the finite correlation time
case, whereas it is $O({\rm St})$ for the zero correlation time model. 

However, a more careful examination of the problem leads to a different conclusion. 
The subtlety lies in considering the statistics of the velocity gradient 
of the particle at its current position, $Y(t)=\partial u/\partial x(x(t),t)$. 
In the case where the correlation time approaches zero, statistics of $Y(t)$ 
are clearly that same as those of $\partial u/\partial x$ sampled at a randomly
chosen point in space. When the correlation time of the velocity field is finite, 
this ergodic relation can be questioned, because the position $x(t)$ has been 
influenced by the history of the velocity field, which is correlated with its 
current value. In section \ref{sec: 4} it was shown that these correlations 
are significant, and imply that the mean value of the velocity gradient at the 
particle position $Y(t)$ is non-zero, despite the fact that the mean value of 
$\partial u/\partial x$ is obviously zero. When this correction
is accounted for, the leading order behaviour of the finite correlation time model
becomes exactly the same as the that of the zero correlation time model, including
the prefactor. 

The principal motivation for this study was to develop a framework for extending 
the analysis of the dynamics of small particles in a random flow to models with 
finite correlation time. The approach developed in the companion paper and in section
\ref{sec: 3} of this work is well suited to this end. The result in section \ref{sec: 4}
indicate that the statistics of the velocity gradient field as sampled by the particles
have to be modelled carefully in order to obtain precise results. 
These conclusions do not contradict the numerical results on the fractal dimension 
presented in \cite{Bec03}, because the calculation of the fractal dimension is
a different problem from that of the largest Lyapunov dimension, considered here.
But they do indicate that the extension to modelling physically realistic flows 
in two and three dimensions will be a delicate issue.

{\sl Acknowledgement}. I have benefitted from extensive discussions with Bernhard
Mehlig, and the calculation in section \ref{sec: 4} was prompted
by his numerical experiments giving evidence for the result reported
there. I am also grateful to Michael Morgan for helpful comments on this paper.

\end{document}